**Title:** Neural circuit function redundancy in brain disorders

**Authors:** Beatriz E.P. Mizusaki[1] and Cian O'Donnell[1]*

[1]Computational Neuroscience Unit, School of Computer Science, Electrical and Electronic Engineering, and Engineering Mathematics, University of Bristol, BS8 1UB, United Kingdom

fv18192@bristol.ac.uk (B.E.P.M.), cian.odonnell@bristol.ac.uk (C.O.)

*corresponding author

**Abstract**

Redundancy is a ubiquitous property of the nervous system. This means that vastly different configurations of cellular and synaptic components can enable the same neural circuit functions. However, until recently very little brain disorder research considered the implications of this characteristic when designing experiments or interpreting data. Here, we first summarise the evidence for redundancy in healthy brains, explaining redundancy and three of its sub-concepts: sloppiness, dependencies, and multiple solutions. We then lay out key implications for brain disorder research, covering recent examples of redundancy effects in experimental studies on psychiatric disorders. Finally, we give predictions for future experiments based on these concepts.

**Redundancy in the nervous system**

Given the astronomically large space of all possible configurations of molecular, cellular and synaptic components, a neural circuit must be constructed in a very particular way to be able to perform useful computations. The task is made easier by the ubiquitous phenomenon of redundancy, which is the idea that, within this enormous space of all possible cellular component configurations there exists a large subset that achieves effectively equivalent macroscopic computations [1,2]. The main empirical evidence for redundancy in neural systems comes from a series of classic papers from Eve Marder and colleagues on a small neural circuit from the crab and lobster stomatogastric ganglia (STG) [1,3]. Using computational models they found that very different arrangements of each STG neuron's ion channels and synaptic conductances could achieve identically sequenced circuit oscillations [4,5]. Accordingly, in experiments these neurons showed 2-3 fold heterogeneity in cellular properties across animals, despite exhibiting consistent circuit function [6]. Similar redundancy phenomena have also been described in Hodgkin-Huxley models [7], mammalian pyramidal neuron models [8], tadpole neurons [9], rodent neuronal activity *in vitro* and *in vivo* [10,11], and human fMRI data [12]. Collectively these studies, plus theoretical arguments [2,13–15], suggest that redundancy is a universal property of the nervous system.

In addition to the core idea of redundancy, we describe three further sub-concepts: sloppiness, compensation, and multiple solutions. *Sloppiness* is the idea that high-level circuit properties are not equally sensitive to the properties of each of its components. Perturbations to some of these components may result in extreme changes to overall function, whereas others may even vary widely while incurring little effect at the circuit level. *Dependence* is a developmental phenomenon where multiple circuit parts are co-tuned with each other, with strong dependencies between their effects on overall function. *Multiple solutions* is the observation that the various configurations of cellular components that enable satisfactory circuit-level functions need not be connected with each other: multiple functional islands can co-exist in the parameter space.

Despite the ubiquity of redundancy in the brain, surprisingly little research on brain disorders has considered its implications when designing experiments or interpreting data. In the remainder of this review we will elaborate these implications and outline how they can be used to guide future brain disorder research.

---

**Box 1. Convergence of brain dysfunction at the level of neural circuits**

Recent high-powered genetic studies have uncovered myriad mutations that correlate with statistical risk for neurodevelopmental and psychiatric disorders [16]. For example, ~100 distinct genetic mutations have been found that elevate risk for Schizophrenia [17], as well as another ~100 that increase risk for Autism Spectrum Disorder (ASD) [18], with overlapping risk gene sets across different psychiatric disorders [19,20]. Despite this heterogeneity at the genetic level, patients present with overlapping symptoms at the cognitive level, and so receive the same umbrella diagnoses. This implies that there must be points of phenotypic convergence within the levels of organisation in the nervous system, which span from molecules to cells, circuits, cognition and behaviour. Neural circuits are a promising focus for analysis for two reasons: first, if molecular, synaptic, or cellular-level alterations in a brain disorder do not lead to alterations in neural circuit function, then they cannot be contributing to cognitive symptoms. Second, because neural circuits are closer to behaviour than cellular components are, it may prove easier to understand or predict the effect of a treatment on behavioural and cognitive symptoms if it is designed to correct dysfunctions at circuit level rather than at molecular level. This argues that in a symptom-targeted approach, we should bias our efforts towards understanding, diagnosing, and treating brain disorders at the neural circuit level rather than the cellular or molecular level, as is common in drug development today [21–24].

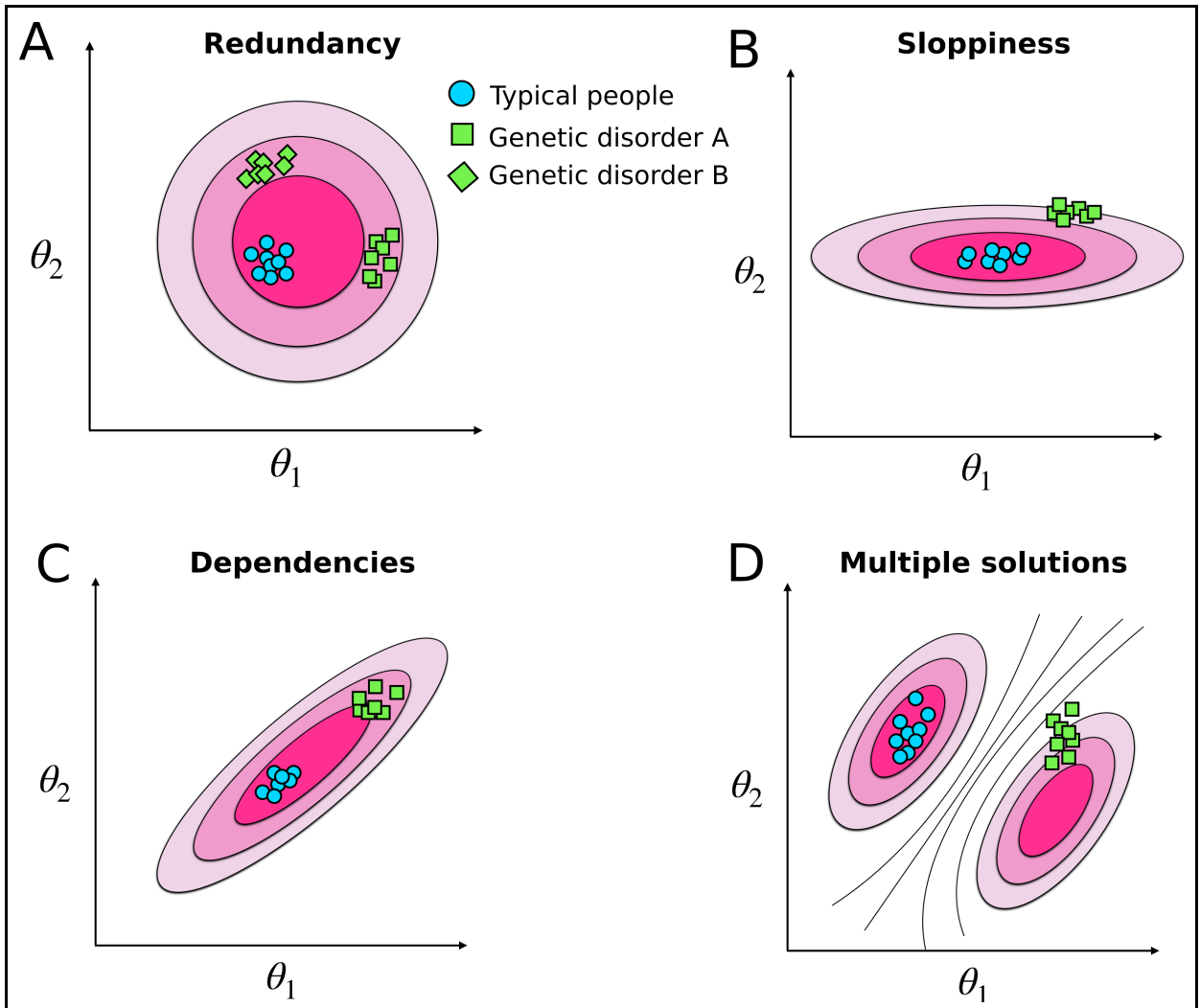

**Figure 1. Forms of redundancy**
The performance of a hypothetical neural circuit is shown as a contour map in pink as a function of the values of two of its components, $\theta_1$ and $\theta_2$. Darker hues of pink represent better circuit performance. Symbols show possible measured values of $\theta_1$ and $\theta_2$ for typical people (blue circles), and two different genetic brain disorders A and B (green squares and diamonds, respectively). Panels A-D show different versions of the contour map illustrating various forms of redundancy: **(A)** generic redundancy, **(B)** sloppiness, **(C)** dependencies, and **(D)** multiple solutions.

**How redundancy is seen and its effects on brain disorder research**

The phenomenon of redundancy and each of its three sub-phenomena (sloppiness, dependencies, and multiple solutions) have distinct implications for brain disorders (Figure 1). First, we illustrate the effects of redundancy itself (Figure 1A) through a measure of the performance of a hypothetical circuit's function shown in a contour plot relative to component parameters, $\theta_1$ and $\theta_2$. Real systems contain thousands of key components, so the parameter space would be much higher-dimensional: our 2D plot is an oversimplification to aid visualization. Optimal circuit function yields high values. Because of evolutionary pressure, we can assume measurements from wild-type animals or neurotypical people will be located near this peak (blue circles) [25]. As an example, two different genetic mutations linked to the same brain disorder may lead to changes in both parameters, drifting affected individuals to different points in the parameter space. Although each genetic mutation may shift the mean parameter changes in a different direction away from the typical case, redundant disorders end up on roughly the same contour line with respect to circuit function. This implies that from the circuit point of view these distinct mutations manifest with the same phenotype, even if their parameters differ.

Importantly however, despite their similarity in circuit function, the two clusters of individuals with brain disorders might be differentially susceptible to perturbations. In the example of the crustacean STG, individual animals may have distinctive sensitivities to changes in temperature, pH, or neuromodulators [26–28]. In our example (Figure 1A), we can imagine some environmental effector such as a drug or stressful life event that causes a small increase in $\theta_1$, corresponding to a rightward shift in all the data points. For both typical people and those with genetic disorder A, this effect would be benign as it would not cause a change in circuit function. In contrast, the same effector could push those with mutation B into even worse values. Alternatively, a different effector that increased $\theta_2$ would not cause a circuit function change in either typical people or those with genetic disorder B, but would have a deleterious effect on those with mutation A. This also illustrates a phenomenon with proposed treatments - they may work to rescue symptoms in one group of patients but not another, even if both groups appear superficially similar. In this sense, redundancy might not only be hiding latent vulnerabilities in the system, but also hiding heterogeneities in those vulnerabilities across patient groups.

Second, molecular or cellular alterations observed in tissue from human patients or animal models may not actually be affecting the circuit-level function – they may be benign. The circuit may be robust to changes in these components over some tolerable range. This property is referred to by different names, according to the research field or author. We will refer to it here as sloppiness [29,30]. Within the same schematic as before, sloppiness can be seen on another hypothetical contour plot (Figure 1B). In this case the circuit function is relatively insensitive to the exact value of one parameter ($\theta_1$), so it may vary horizontally in the plot across a large range without causing much change in circuit function. In contrast, small changes to the other parameter ($\theta_2$) will induce large changes in circuit function. In this case $\theta_1$ is the sloppy parameter. If we consider a brain disorder where a genetic mutation tends to increase both parameters $\theta_1$ and $\theta_2$ in the brains of affected individuals, the change in $\theta_2$ would be the primary driver of dysfunction, although $\theta_1$'s value would still be correlated with disease severity. If an experimental scientist measured the value of $\theta_1$ in both wild-type and brain disorder animal models, they may see a clear difference in the group mean values of $\theta_1$, and a parallel change in circuit function. They may conclude that the changes in $\theta_1$ are responsible for the circuit level-deficits, and design an intervention to reverse the molecular level change in $\theta_1$; however, the treatment would not be successful.

Third, we comment on redundant dependencies: altered components may individually have large effects on circuit function when perturbed genetically or experimentally, but homeostatic processes during development may restore high-level function by compensating with changes in other circuit components. In a simple case this could be a straight pairing of opposing factors, such as increased expression of sodium channels that depolarize the cell being counteracted by increased expression of potassium channels that hyperpolarize it. However, in intact brains there are so many nonlinear interactions that the compensatory relationships might not be obvious from raw measurements. In Figure 1C we depict this idea with another hypothetical contour map on a 2D parameter space. In this case proper circuit function requires jointly low or jointly high values of $\theta_1$ and $\theta_2$ together, so if one parameter is low while the other is high then circuit function is impaired. A genetic mutation could cause a direct increase in $\theta_1$, but be developmentally compensated by a corresponding increase in $\theta_2$. In this situation, an experiment may yield clear group-level differences in $\theta_1$ between wild-type and animal models, but they may not measure parallel changes in $\theta_2$. If a scientist then goes on to design an intervention to bring the value of $\theta_1$ in the animal model back down to wild-type values without altering $\theta_2$, they might inadvertently make the circuit function worse, not better.

Last, there may be multiple distinct solutions to the circuit design (Figure 1D), appearing as multiple islands. Although it is likely that these peaks may be connected via some paths in the full high-dimensional parameter space of all circuit components [31,32], any experimental measurement of a small subset of parameters or therapeutic intervention may access only a low-dimensional subspace, where such local optima are likely to persist. Even though the phenomenon of multiple solutions complicates our attempts to understand how brains work, it could paradoxically end up simplifying our search for brain disorder interventions. It implies that fixing circuit function does not require a straight reversal of the original alteration. Depending on how many solutions exist, it may instead be more practical to find a new configuration that restores the circuit operating mode, rather than trying to undo all the various component changes that have accumulated across development – most of which are in any case likely to be hidden to the experimentalist or clinician. We illustrate the multiple solutions phenomenon schematically in Figure 1D. The nearest part of parameter space that rescues circuit function in the brain disorder case is not the same as for the neurotypical case. Also, the intervention that implements this correction would involve changing only $\theta_2$, even though $\theta_1$ was the parameter altered by the original genetic mutation. Therefore, the phenomenon of multiple solutions may open up counter-intuitive options for therapeutics.

**Example of redundancy in brain disorders**

Although few studies have directly explored the consequences of redundancy in brain disorders, many have found evidence for homeostatic compensation where changes in one brain component seemed to be counterbalanced by changes in others, a form of redundancy [33–38]. There is also evidence for disrupted homeostatic plasticity [39–41] and proposals for how global brain perturbations could lead to deficits only in select neural circuits [42]. However, one recent study by Antoine et al [43] found explicit evidence for circuit redundancy in mouse models of autism. The authors used patch-clamp electrophysiology to measure excitatory and inhibitory synaptic inputs from layer 4 onto single layer 2/3 pyramidal neurons in brain slices of primary somatosensory cortex from wild-type mice, and from four genetic mouse models of autism. Nominally their aim was to ask if the ratio of synaptic excitation to inhibition (E/I balance) was altered in the autism mouse models, a common theory for autism [44,45]. Indeed they found that in each of the four autism models inhibition was decreased more than excitation, implying an increase in the E/I ratio compared to wild-type mice. Surprisingly however, they also found that in each case the amplitude of postsynaptic potentials (PSP) or spiking responses to stimulation was unchanged relative to wild-type controls. The authors explained this mismatch via computational modelling, which showed that a range of different synaptic E/I ratios would be consistent with any given PSP amplitude (Figure 2). Although the genetic mutations were causing real shifts in synaptic properties, their effects were redundant at the level of the neuron's net response to synaptic stimulation. The authors speculated that this may have been achieved by compensatory homeostatic mechanisms during development [46]. Overall the result suggested that the autism field's decade-long search for E/I imbalance may have been misguided, since redundancy nullified its apparent effect on circuit function.

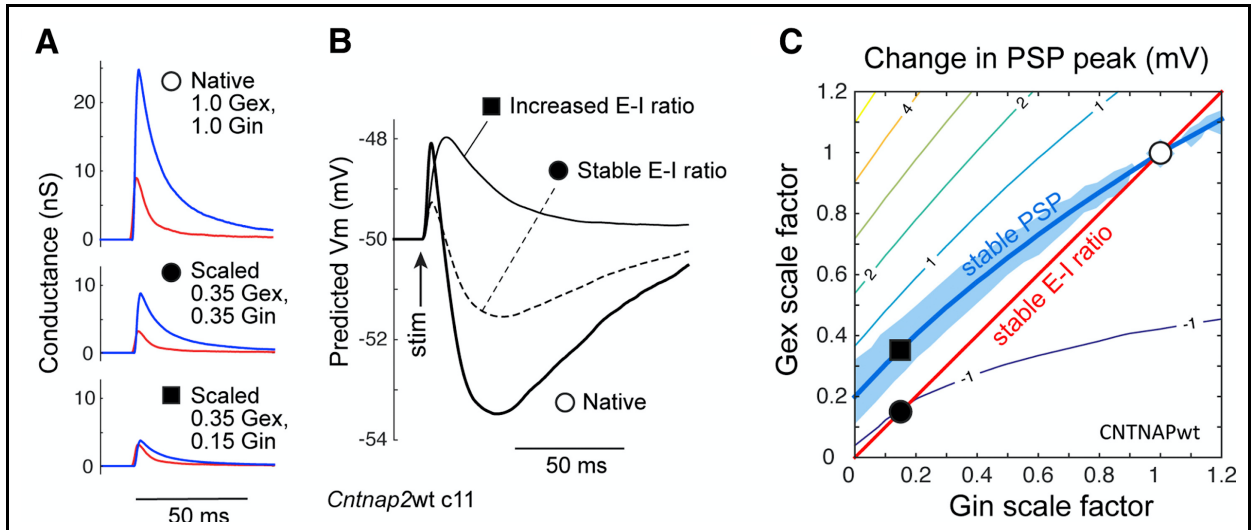

**Figure 2. Redundancy in mouse models of autism.**
A: Example excitatory (red) and inhibitory (blue) synaptic conductance time series from a basic computational model of somatic voltage. Top "Native" plot shows when synaptic conductances are set to the values estimated from layer 4 to layer 2/3 synapses in wild-type mice. In *Cntnap2* knock-out animals, a model for autism, excitatory and inhibitory synaptic conductances were decreased to 35% and 15% of wild-type values, respectively, implying an increase in the excitation/inhibition ratio. Middle plot shows traces if both conductances were scaled equally to 35% of wild-type values, bottom plot shows situation to match the data, where inhibition is decreased more than excitation.
B: Compound postsynaptic potentials (PSPs) corresponding to the three scenarios shown in panel A. Note that PSP amplitude is decreased relative to Native case if E-I ratio is kept fixed, whereas the increased E-I ratio keeps PSP amplitude matched to Native.
C: Contour map of peak PSP amplitude as a function of scaling factor on excitatory and inhibitory synaptic strengths. Open circle is mean from wild-type control animals. Red line corresponds to fixed E-I ratio, blue line corresponds to fixed PSP peak. Black square symbol is mean value of synaptic strengths in *Cntnap2* knock-out mice, while black circle symbol is where values would lie if the E-I ratio was stable.
Figure adapted with permission from [43].

Another recent study, by O'Donnell et al [47], found using a computational model of the same brain region, mouse L2/3 somatosensory cortex, that circuit-level function shows extreme differences in sensitivity to perturbations in some components over others, corresponding to sloppiness (Figure 1B). In line with previous studies [48], the authors also found that neural correlations were altered in a mouse model of Fragile-X syndrome, but this circuit-function-level property did not map neatly onto any one distinct circuit model component, implying both redundancies and dependencies between parameters (as in Figure 1C).

**Conclusion**

In summary, because redundancy appears to be a ubiquitous feature of the nervous system, we argue that it should be a central consideration when trying to understand or develop treatments for brain disorders. We also believe it offers hope, as the rapid pace of the field of neuroscience of psychiatric disorders can sometimes make us feel as though we are lost in complexity, that our collective understanding is becoming murkier as research progresses, not clearer. A tactful consideration of redundancy may offer us a way through, because it means that we may not need to classify and measure every last detail of every form of disorder in order to develop effective treatments. Instead, we may be able to discover generic principles of neural circuit dysfunction that allow us to generalize our insights across molecularly distinct disorders, and develop rational treatment strategies that enable correction of systems-level symptoms.

---

**Box 2. Predictions**

As this is a very general framework we anticipate that many predictions follow. Here we give one example prediction for redundancy and each of its three sub-phenomena.

- *Redundancy* predicts that the magnitude of the differences in measures of neural circuit components between genotypes is greater than the magnitude of differences of measures of functional activity in the same circuits. However, this superficial similarity may hide heterogeneity in response to perturbations, across groups of related disorders.
- *Sloppiness* predicts that the degree of within- or across-animal heterogeneity in a circuit component parameter should be inversely proportional to the magnitude of its effect on circuit function. If a particular component shows low heterogeneity across wild-type animals, and it is altered in a brain disorder, then it likely also plays a causal role in any circuit-function level alterations.
- *Dependence* predicts that any set of cellular components that strongly co-vary within wild-type animals are unlikely to be causally contributing to circuit-function level alterations in brain disorders. Reversing the changes of any subset of these components in isolation may exacerbate circuit-function symptoms.

> - *Multiple solutions* predicts that the individual animals from a genetically modified cohort that are most similar to the wild-type animals at the circuit function or behavioral level will not necessarily have the most wild-type-like circuit components.

**Acknowledgements**

We would like to thank Alan Anticevic, John Murray, and Timothy O'Leary for helpful discussions. This work was supported by funding from the Simons Foundation Autism Research Initiative (RPG-2019-229) and the Medical Research Council (MR/S026630/1).

**Declarations of interest:** none.